\documentclass[10pt,a4paper,showkeys,showpacs,twocolumn]{revtex4-1}

\usepackage{url}
\usepackage[english]{babel}
\usepackage{times}
\usepackage{graphics}
\usepackage{epsfig}
\usepackage{pstricks}
\usepackage{xcolor}
\usepackage{amssymb,amsmath,color}

\newcommand{\ket}[1]{{| {#1} \rangle}}

\newcommand{\braket}[2]{{\langle {#1} |{#2}\rangle}}

\newcommand{\reffig}[1]{\figurename{~\ref{#1}}}
\newcommand{\refapp}[1]{Appendix~\ref{#1}}
\newcommand{\etal}{\textit{et al}}

\begin{document}


\title{{\color{red}{Laser Physics PROOFS}}\\Analysis of elliptically polarized maximally entangled states for Bell inequality tests}

\author{A. Martin$^{1}$, J.-L. Smirr$^{2}$,  F. Kaiser$^{1}$, E. Diamanti$^{2}$, A. Issautier$^{1}$,\\
O. Alibart$^{1}$, R. Frey$^{2}$, I. Zaquine$^{2}$, and S. Tanzilli$^{1}$\\
$^1$Laboratoire de Physique de la Mati\`ere Condens\'ee, CNRS UMR 7336,\\
Universit\'e de Nice - Sophia Antipolis, Parc Valrose, 06108 Nice Cedex 2, France\\
$^2$Laboratoire Traitement et Communication de l'Information, CNRS UMR 5141,\\
Institut Mines T\'el\'ecom/T\'el\'ecom ParisTech, 46 rue Barrault, 75013 Paris, France\\
e-mail: sebastien.tanzilli@unice.fr; isabelle.zaquine@telecom-paristech.fr
}

\begin{abstract}When elliptically polarized maximally entangled states are considered, \textit{i.e.}, states having a non random phase factor between the two bipartite polarization components, the standard settings used for optimal violation of Bell inequalities are no longer adapted. One way to retrieve the maximal amount of violation is to compensate for this phase while keeping the standard Bell inequality analysis settings. We propose in this paper a general theoretical approach that allows determining and adjusting the phase of elliptically polarized maximally entangled states in order to optimize the violation of Bell inequalities. The formalism is also applied to several suggested experimental phase compensation schemes. In order to emphasize the simplicity and relevance of our approach, we also describe an experimental implementation using a standard Soleil-Babinet phase compensator. This device is employed to correct the phase that appears in the maximally entangled state generated from a type-II nonlinear photon-pair source after the photons are created and distributed over fiber channels.
\end{abstract}



\maketitle

\textbf{ }

\newpage

\textbf{ }

\newpage

\section{Introduction}

Entanglement is an essential resource for many quantum information protocols, such as quantum key distribution~\cite{scarani_quantum_2009}, quantum teleportation~\cite{kim_quantum_2001}, entanglement swapping~\cite{halder_entangling_2007}, quantum relays~\cite{Collins_quantum_2005,Aboussouan_pico_2010}, quantum memories and repeaters~\cite{briegel_1998,duan_2001,simon_2007,Hammerer_2010}, as well as for quantum algorithms~\cite{politi_shors_2009,obrien_photonic_2009}. Determining the potential amount of entanglement delivered by a source, or by any quantum information system, is therefore of prime interest.
Historically, the first entanglement witness was introduced by Bell in 1964 as an inequality~\cite{bell_1964}, which was reformulated a few years later by Clauser, Horne, Shimony, and Holt (CHSH)~\cite{CHSH_1969}. Since then, more witnesses have been reported in the literature for further investigation of entanglement~\cite{altepeter_experimental_2005}, \textit{i.e.}, dedicated to states that are non pure and/or non maximally entangled. Among others, inequalities proposed by Yu \etal~\cite{Yu_Pan_Ineq_2003}, and by Collins and Gisin~\cite{CG_Ineq_2004}, are now widely employed experimentally. In addition to these tests, a complete way to analyze quantum states is known as quantum state tomography~\cite{altepeter_experimental_2005,james_measurement_2001}. This method is complete in the sense that it permits reconstructing the full density operator of the quantum state under test (although it does not constitute a non-locality test). Such a method requires, however, a minimum of 16 measurements to reconstruct the density operator matrix. Depending on the source under test, these measurements can take a long time before providing the state parameters, contrary to the standard CHSH inequality measurement. In the following analysis, we will thus be interested in Bell-type inequality tests.

\begin{figure}[h]
\resizebox{0.75\columnwidth}{!}{\includegraphics{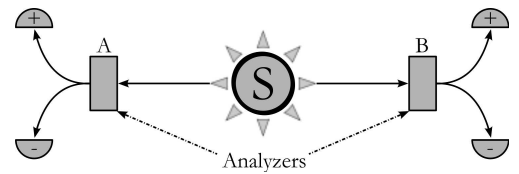}}
\caption{\label{Fig:1}Schematics of the considered setup. A source (S) produces pairs of maximally polarization entangled photons. Rotating polarization analyzers (\textit{e.g.}, polarizing beamsplitters) are placed on the path of each photon, \textit{i.e.}, on channels $A$ and $B$, respectively. The associated rotations are defined by an angle $i=a,b$, with respect to the $\{\ket{H_A},\ket{V_A}\}$, $\{\ket{H_B},\ket{V_B}\}$ basis, respectively.}
\end{figure}

Let us consider the setup depicted in \reffig{Fig:1}, which corresponds to a Bell inequality test of polarization entangled photons. Using the basis $\{\ket{H_I},\ket{V_I}\}$, referring to horizontal and vertical polarizations on channel $I$, for $I=\{A,B\}$, any polarization entangled state, and more generally two-system state, can be decomposed into the basis $\lbrace\ket{H_A,H_B}$,  $\ket{H_A,V_B}$, $\ket{V_A,H_B}$, $\ket{V_A,V_B}\rbrace$. Including a phase factor $\phi$ between the two components of the state, we describe bipartite, elliptically polarized, maximally entangled states as follows:
\begin{equation}\label{Eq:Phi}
\begin{array}{l c l}
\ket{\Phi(\phi)}&=&\frac{1}{\sqrt{2}}\left[\ket{H_A,H_B}+e^{i\phi}\ket{V_A,V_B}\right]\\
\ket{\Psi(\phi)}&=&\frac{1}{\sqrt{2}}\left[\ket{H_A,V_B}+e^{i\phi}\ket{V_A,H_B}\right].
\end{array}
\end{equation}
In photonic setups, an arbitrary, but non random, phase, as in Eq.~(\ref{Eq:Phi}), is very common and can either be induced by the source itself~\cite{kwiat_typeII_1995,kwiat_ultrabright_1999,quantum_dot_Mohan,quantum_dot_Beveratos}, or be accumulated over the quantum channel linking the source to the analyzers, especially when this consists of optical fibers~\cite{Martin_NJP_2010}. The Bell inequality violation is sensitive to this phase factor. Indeed, its existence does not allow for the optimal violation of the CHSH inequality when the standard settings are used for the polarization analyzers~\cite{CHSH_1969}. One way to retrieve the maximal amount of violation is to compensate for this phase while keeping the nominal settings~\cite{kwiat_typeII_1995,Martin_NJP_2010}. Consequently, determining the phase experimentally makes it possible to adapt the state towards optimal entanglement measurements.

The described scenario is typically encountered in experimental setups of Bell inequality tests for polarization entangled states. Phase measurement and compensation techniques are often empirically used to recover the optimal violation; here we provide a mathematical formalization of the problem, which allows comparing the various techniques and adapt the measurement settings to the states under test in various experimental configurations. Note that the $\ket{\Psi(\phi)}$ state in Eq.~(\ref{Eq:Phi}) can directly be derived from the $\ket{\Phi(\phi)}$ state through a $\pi/2$ rotation of the polarization on channel $B$, thus we will restrict the following analysis to the state $\ket{\Phi(\phi)}$.



The paper is organized as follows.
Section~\ref{Sec:PS} is devoted to the standard analysis of elliptically polarized maximally entangled states $\ket{\Phi(\phi)}$, using standard settings and rotating polarization analyzers, recalling that the maximum value $|S|=2\sqrt{2}$ of the Bell parameter can only be obtained for the $\ket{\Phi_+}=\ket{\Phi(0)}$ and $\ket{\Phi_-}=\ket{\Phi(\pi)}$ states. Section~\ref{Sec:GF} presents the general formalism developed to derive the necessary conditions towards maximizing the Bell parameter for any state of the form $\ket{\Phi(\phi)}$. The next two sections present theoretical implementations of this approach using potential experimental configurations. More precisely, Section~\ref{Sec:DM} deals with one rotating phase compensator, while Section~\ref{Sec:piezo} considers a set of two phase compensators at fixed angles. Finally, in Section~\ref{Sec:exp} we demonstrate an experimental implementation, based on a type-II non-linear photon-pair source, in which the phase of the state is adjusted after the photons are created and distributed over fiber channels.

\section{Analysis with rotating polarization analyzers}\label{Sec:PS}

In the following, we consider the standard Bell inequality test setup shown in \reffig{Fig:1}, and we perform the standard calculation leading to the maximum Bell parameter value using optimized analyzer settings for a given phase.

Any elliptically polarized Bell state $\ket{\Phi(\phi)}$ can be written as a superposition of $\ket{\Phi_+}$ and $\ket{\Phi_-}$, \textit{i.e.}, as $\ket{\Phi(\phi)}=\frac{1+e^{i\phi}}{2} \ket{\Phi_+}+\frac{1-e^{i\phi}}{2} \ket{\Phi_-}$.
We define the basis $\{\ket{+_I},\ket{-_I}\}$, where $I=\{A,B\}$, related to the initial basis $\{\ket{H_I},\ket{V_I}\}$ using the transformation
\begin{equation}\label{Eq:transform}
\begin{pmatrix}
\ket{+_I}\\\ket{-_I}
\end{pmatrix}
=\begin{pmatrix}
\cos \theta_I &\sin \theta_I\\
-\sin \theta_I &\cos \theta_I
\end{pmatrix}
\begin{pmatrix}
\ket{H_I}\\\ket{V_I}
\end{pmatrix},
\end{equation}
where $\theta_I = a, b $ is the angle between the analyzer $I$ and $\ket{H_I}$, for $I=\{A,B\}$, respectively.

The outcome probabilities, $P_{++}^{(\phi)}(a,b)=|\braket{+_A+_B}{\Phi(\phi)}|^2$ and $P_{--}^{(\phi)}(a,b)=|\braket{-_A-_B}{\Phi(\phi)}|^2$, of measuring both photons in the same state are given by
\begin{equation}\label{Eq:P++ab}
\begin{array}{l c l}
 P_{++}^{(\phi)}(a,b)&=&P_{--}^{(\phi)}(a,b)\\
 &=&\frac{1}{2}\left[\cos^2(\frac{\phi}{2})\cos^2(a-b)\right.\\
 &&\left. +\sin^2(\frac{\phi}{2})\cos^2(a+b)\right],
\end{array}
\end{equation}
while the outcome probabilities $P_{+-}^{(\phi)}(a,b)=|\braket{+_A-_B}{\Phi(\phi)}|^2$ and $P_{-+}^{(\phi)}(a,b)=|\braket{-_A+_B}{\Phi(\phi)}|^2$ of measuring both photons in different states read
\begin{equation}\label{Eq:P+-ab}
\begin{array}{l c l}
 P_{+-}^{(\phi)}(a,b)&=&P_{-+}^{(\phi)}(a,b)\\
 &=&\frac{1}{2}\left[\cos^2(\phi/2)\sin^2(a-b)\right.\\
 &&\left. +\sin^2(\phi/2)\sin^2(a+b)\right].
\end{array}
\end{equation}
Note that changing both $\phi$ to $\phi +\pi$ and $b$ to $-b$ leaves Eqs.~(\ref{Eq:P++ab}) and (\ref{Eq:P+-ab}) unchanged, so that we can restrict the analysis to $0 \leq \phi \leq \pi$.

The correlation factor $E^{(\phi)}(a,b)=P_{++}^{(\phi)}(a,b)-P_{+-}^{(\phi)}(a,b)-P_{-+}^{(\phi)}(a,b)+P_{--}^{(\phi)}(a,b)$, as defined in Ref.~\cite{bell_1964}, is then given by
\begin{equation}
E^{(\phi)}(a,b)=\cos^2(\phi/2)\cos 2(a-b)+\sin^2(\phi/2)\cos2(a+b).
\end{equation}
The Bell parameter, defined as
\begin{equation}\label{Eq:S}
\begin{array}{l c l}
S^{(\phi)}(a,b,a',b')&=&E^{(\phi)}(a,b)+E^{(\phi)}(a,b')\\
	&&+E^{(\phi)}(a',b)-E^{(\phi)}(a',b'),
\end{array}
\end{equation}
is therefore given by
\begin{equation}\label{Eq:bell_param}
\begin{array}{r}
 S^{(\phi)}(a,b,a',b')=\cos^2(\frac{\phi}{2})\left[\cos 2(a-b)+\cos 2(a-b')\right.\\
 +\left.\cos 2(a'-b)-\cos 2(a'-b')\right]\\
	 +\sin^2(\frac{\phi}{2})\left[\cos 2(a+b)+\cos 2(a+b')\right.\\
	 \left.+\cos 2(a'+b)-\cos 2(a'+b')\right].
\end{array}
\end{equation}

To demonstrate a violation of the Bell inequality, \textit{i.e.}, $|S^{(\phi)}(a,b,a',b')| >2$~\cite{bell_1964,CHSH_1969}, it is necessary to choose the analysis parameters $a$, $a'$, $b$, and $b'$ so as to maximize $S^{(\phi)}(a,b,a',b')$ for any given value of $\phi$. The optimal angles are obtained by solving the set of equations $\left( \frac{\partial S^{(\phi)}}{\partial a}=0,\frac{\partial S^{(\phi)}}{\partial b}=0,\frac{\partial S^{(\phi)}}{\partial a'}=0,\frac{\partial S^{(\phi)}}{\partial b'}=0\right)$.\\
Using trigonometric transformations, this set of equations can be written in the simple form:
\begin{equation}
\left \{ \begin{array}{l c l}
\tan 2a&=&\cos \phi \,\tan (b+b')\\
\tan 2b&=&\cos \phi \, \tan (a+a')\\
\tan 2a'&=&- \cos \phi \, \mathrm{cotan} (b+b')\\
\tan 2b'&=&-\cos \phi\, \mathrm{cotan} (a+a').
\end{array}
\right.
\end{equation}
Solving these equations gives the solutions $a=0$, $a'=\pi/4$, and $b=-b'=\frac{1}{2}\arctan (\cos \phi)$, and using Eq.~(\ref{Eq:bell_param}), we obtain
\begin{equation}
\begin{array}{lcl}
 S^{(\phi)}_{Max}&=&2\left[\cos (\arctan(\cos \phi))\right.\\
 	&&\left.+\cos \phi \sin(\arctan(\cos \phi))\right]\\
 &=&2\sqrt{\cos^2(\phi)+1}.
\end{array}
\end{equation}
\begin{figure}
\resizebox{1\columnwidth}{!}{\includegraphics{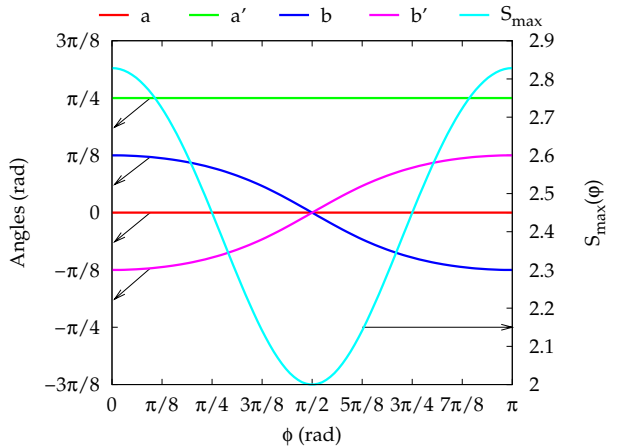}}
\caption{\label{Fig:2}Maximum values of the Bell parameter $S^{(\phi)}$ for a set of setting ($a$, $a'$, $b$, $b'$) ploted as a function of $\phi$ for standard calculation and optimization.}
\end{figure}
The rotation angles of the analyzers ($a$, $a'$, $b$, $b'$) and $S^{(\phi)}_{Max}$ are plotted as a function of $\phi$ in \reffig{Fig:2}. We find, as expected, the usual set of parameters for phase values equal to 0 and $\pi$, which leads to a maximal violation of the Bell inequality. Moreover, the Bell inequality is violated ($S^{(\phi)}_{Max}>2$) for any phase except for the value $\phi=\pi/2$, for which $S _{Max}= 2$. This corresponds to the circularly polarized maximally entangled state $\ket{\Phi(\pi/2)}= \frac{1}{\sqrt{2}}\left[\ket{H_AH_B} + i\ket{V_AV_B}\right]$. In practical systems, however, the measurement precision is always limited by noise, which extends to a larger range of $\phi$ values around $\pi/2$ the domain where the Bell inequality is not violated. Rotating analyzers are therefore unsuitable for optimal entanglement measurements if the phase $\phi$ of the state is not determined and compensated for by an appropriate polarization device.

We develop in the following a formalism for recovering optimal settings for Bell inequality violation by any elliptically polarized maximally entangled state, and apply it to two practical phase compensation configurations (Secs. \ref{Sec:DM} and \ref{Sec:piezo}). We finally give an actual experimental situation in which our formalism is directly implemented (Sec. \ref{Sec:exp}).

\section{General formalism for optimal analysis of elliptically polarized, maximally entangled states\label{Sec:GF}}

To retrieve the optimal violation of the Bell inequality for any elliptically polarized, maximally entangled state, polarization devices, labeled $T_A$ and $T_B$, are introduced between the source and the two initially rotating analyzers ($A$ and $B$) depicted in \reffig{Fig:1}. The following analysis takes into account these devices that are considered adjustable, while the two former polarization analyzers are now kept fixed (\textit{e.g.}, standard polarization beamsplitters). The new analysis parameters to be further defined are therefore directly linked to $T_A$ and $T_B$ settings. To generalize the formalism of Eq.~(\ref{Eq:transform}), we define the unitary transformations $\hat{T}_A$ and $\hat{T}_B$ as
\begin{equation}\label{Eq:TI}
\begin{pmatrix}
\ket{+_I}\\
\ket{-_I}
\end{pmatrix}
= \hat{T}_I
\begin{pmatrix}
\ket{H_I}\\
\ket{V_I}
\end{pmatrix}=
\begin{pmatrix}
h_I&v_I\\
-v_I^*&h_I^*
\end{pmatrix}
\begin{pmatrix}
\ket{H_I}\\
\ket{V_I}
\end{pmatrix},
\end{equation}
with $I=\{A,B\}$. $h_I$ and $v_I$, which satisfy the condition $|h_I|^2+|v_I|^2=1$, can be defined as
\begin{equation}
\left \{ \begin{array}{l c l}
 h_I&=&\cos\alpha_I \;e^{i\varphi_I}\label{Eq:hI}\\
 v_I&=&\sin\alpha_I \;e^{i\varphi'_I}
\end{array}
\right.
\end{equation}

In the same way as in Section~\ref{Sec:PS}, we can infer the outcome probabilities $P_{++}^{(\phi)}$, $P_{+-}^{(\phi)}$, $P_{-+}^{(\phi)}$, and $P_{--}^{(\phi)}$, using Eq.~(\ref{Eq:TI}), as follows:
\begin{equation}\label{Eq:P++}
\begin{array}{l c l}
P_{++}^{(\phi)}=P_{--}^{(\phi)} &=&\frac{1}{2}\left[\cos^2\frac{\phi_{eff}}{2}\cos^2(\alpha_A-\alpha_B)\right.\\
&&+\left.\sin^2\frac{\phi_{\text{eff}}}{2}\cos^2(\alpha_A+\alpha_B)\right]\\
P_{+-}^{(\phi)}=P_{-+}^{(\phi)} &=& \frac{1}{2}\left[\cos^2\frac{\phi_{eff}}{2}\sin^2(\alpha_A-\alpha_B)\right.\\
&&+\left.\sin^2\frac{\phi_{\text{eff}}}{2}\sin^2(\alpha_A+\alpha_B)\right],
\end{array}
\end{equation}
where the effective phase, $\phi_{\text{eff}}$, is given by
\begin{equation}\label{Eq:phieff}
\phi_{\text{eff}}=\phi+\varphi_A+\varphi_B-\varphi'_A-\varphi'_B.
\end{equation}
The maximum value of the Bell parameter ($S=2\sqrt{2}$) can obviously be obtained, for instance, for $\phi_{\text{eff}} =0$ (mod $2\pi$). In this case, Eq.~(\ref{Eq:P++}) corresponds exactly to the measurement of the $\ket{\Phi_+}$ state (see Section~\ref{Sec:PS}). However, the analysis parameters are no longer the former analyzer settings $\{a,b\}$ but rather the coefficients $\alpha_A$ and $\alpha_B$, corresponding to $\hat{T}_A$ and $\hat{T}_B$, respectively. Note that the Bell inequality test can be performed using positive and negative angle settings. In the following, we restrict the analysis to the case $\alpha_I \geq 0$. The case $\alpha_I<0$ is directly obtained by changing $\varphi'_I$ into $\varphi'_I + \pi$ in Eq.~(\ref{Eq:hI}).

The phase value $\phi$ can be conveniently determined experimentally from the coincidence measurements in the diagonal basis, \textit{i.e.}, for $\alpha_A = \alpha_B = \pi/4$, since coincidence and anti-coincidence probabilities are then reduced to
\begin{equation}\label{Eq:phaseadj}
\begin{array}{l c l}
P_{++}^{(\phi)}&=& P_{--}^{(\phi)} =\frac{1}{2} \cos^2 \frac{\phi_{\text{eff}}}{2}\\
P_{+-}^{(\phi)}&=& P_{-+}^{(\phi)} =\frac{1}{2} \sin^2 \frac{\phi_{\text{eff}}}{2},
\end{array}
\end{equation}
where $\phi_{\text{eff}}$ is linked to $\phi$ through Eq.~(\ref{Eq:phieff}). We can infer the phase $\phi$ from a fit of the measured probability $P_{++}^{(\phi)}$ as a function of a polarization device dependent parameter linked to $\varphi_{I}$ or $\varphi'_{I}$.Then, for any given transformation $\hat{T}_A,\hat{T}_B$ satisfying $\phi_{\text{eff}}=0$ (Eq.~(\ref{Eq:phieff})), the optimal value $|S|=2\sqrt{2}$ for the Bell parameter can be retrieved using the appropriate settings of the new analysis parameters $\alpha_A,\alpha_B$.

It should be stressed that the transformations $\hat T_A$ and $\hat T_B$ contain six free parameters ($\alpha_A,\,\alpha_B,\,\varphi_A,\,\varphi_B,\, \varphi'_A$ and $\varphi'_B$). The techniques that are detailed hereafter to analyze elliptically polarized states for maximal violation of Bell inequalities, require only three free parameters (one analysis parameter on each channel and one phase parameter to satisfy the relation $\phi_{\text{eff}}=0$ (mod $2\,\pi$)). Therefore, without loss of generality, we will choose to fix some of the parameters in each practical case that we study, depending on the considered device.

\section{Analysis using a rotating phase compensator}\label{Sec:DM}

In this section, we apply the formalism developed above to the case of the most general type of polarization device, \textit{i.e.}, a rotating phase compensator. A free space optics based realization of such a device is a rotating Soleil-Babinet phase compensator. We note that there are also fully fibered solutions, where mechanical stress is applied to the fiber. Then, the amount of stress controls the introduced birefringence and phase between the two polarization modes which are parallel and orthogonal to the stress direction, while the rotation angle parameter is controlled by the direction in which the stress is applied.

To perform our analysis, we represent operators $\hat{T}_A$ and $\hat{T}_B$, quantitatively, by a Jones matrix, as explained in detail in \refapp{A:Jones}. Then, the set of parameters ($h,v$) of the unitary transformation are related to the retardation, $\chi$, and rotation, $\zeta$, angles of the devices on each channel. We consider now a configuration, in which the direct measurement of the Bell parameter $S$ for an elliptically polarized, maximally entangled state is performed with one rotating phase compensator on channel $A$ and a phase compensator set at $\zeta_B=\pi/4$ with respect to the $\{\ket{H_B},\ket{V_B}\}$ basis on channel $B$~\cite{Ekert_Crypto_91}. Then, from Eqs.~(\ref{Eq:hI}) and (\ref{Eq:Tphipi/4}) we obtain $\alpha_B = \chi_B/2$, $\varphi_B=0$, and $\varphi'_B=\pi/2$.

Furthermore, identifying the real and imaginary parts of $h_A$ and $v_A$, based on Eqs.~(\ref{Eq:hI}) and (\ref{Eq:h}), allows determining the coefficients $\alpha_A$, $\varphi_A$, and $\varphi'_A$ of the unitary transformation $\hat{T}_A$:
\begin{equation}\label{Eq:alphaA}
\left \{ \begin{array}{l c l}
\alpha_A&=&\arcsin\left[\sin\frac{\chi_A}{2}\sin2\zeta_A\right] \\
\varphi_A&=&\arctan\left[-\tan\frac{\chi_A}{2}\cos2\zeta_A\right] \\
\varphi'_A&=&-\frac{\pi}{2}.
\end{array}
\right.
\end{equation}
Using these rotating and fixed phase compensators on channels $A$ and $B$, respectively, the condition $\phi_{\text{eff}} = 0$ (mod $2\pi$) simply gives $\varphi_A = -\phi-\pi$. Then,
the Bell inequality measurements can be performed using $\alpha_A$ and $\alpha_B$ as new analysis parameters, whose values can be obtained from 
\begin{equation}\label{Eq:chiA}
\left \{ \begin{array}{l c l}
\chi_B&=&2\alpha_B\\
\chi_A&=&2\arccos\left[\cos\alpha_A\cos\phi\right] \\
\zeta_A&=&\frac{1}{2}\arcsin\left[\frac{\sin\alpha_A}{\sin\frac{\chi_A}{2}}\right].
\end{array}
\right.
\end{equation}



The phase $\phi$ must be determined before making entanglement measurements. The condition $\alpha_A=\alpha_B=\pi/4$, used to derive Eq.~(\ref{Eq:phaseadj}), leads to
\begin{equation}\label{Eq:22}
\left \{ \begin{array}{l c l}
\chi_B&=&\pi/2\\
\zeta_A &=& \frac{1}{2}\arcsin\frac{1}{\sqrt{2}\sin\chi_A/2}.
\end{array}
\right.
\end{equation}
By inserting $\phi_{\text{eff}}=\phi+\varphi_A$ in Eq.~(\ref{Eq:phaseadj}), we obtain:
\begin{equation}\label{Eq:P++phiA}
\begin{array}{l c l}
P_{++}^{(\phi)}(\varphi_A)&=&\frac{1}{2}\cos^2\left(\frac{\phi+\varphi_A}{2}\right),
\end{array}
\end{equation}
with $\varphi_A=\arccos(\sqrt{2}\cos\frac{\chi_A}{2})$. Phase $\phi$ can therefore be obtained from the measurement of $P_{++}^{(\phi)}$ as a function of $\varphi_A$. Then, the
retardation angle $\chi_A$ can be chosen in order to satisfy $\phi_{\text{eff}}=0$ and we are left with the free parameters $\chi_B$ and $\zeta_A$ for the Bell inequality measurement.

\section{Analysis using a set of phase compensators at fixed angle}\label{Sec:piezo}

In the case of fiber quantum communication systems, if the phase of the elliptically polarized state is due to the polarization fluctuations induced by the network optical fiber, a
real time adjustment is necessary. It is possible in this case to use commercially available phase compensators composed of effective fiber waveplates, for which birefringence is modified along two given directions through a stress induced by a piezo-electric actuator. These devices are the fiber equivalent to the standard bulk-optics Soleil-Babinet compensators and the voltages applied to the piezo-electric actuators can be easily controlled dynamically. As we will show in the following, such a setup allows a convenient estimation of $\phi$ and measurement of the Bell parameter.

Following previously reported experimental results~\cite{kwiat_ultrabright_1999,kwiat_typeII_1995,Martin_NJP_2010}, we apply the general formalism presented in Section~\ref{Sec:GF} to
a configuration, in which there is one phase compensator set at $\zeta_B=\pi/4$ with respect to the $\{\ket{H_B},\ket{V_B}\}$ basis on channel $B$ and a set of two compensators
(the first set parallel to the $\{\ket{H_A},\ket{V_A}\}$ basis ($\zeta_{1A}=0$) and the second at $\zeta_{2A}=\pi/4$ with respect to the same basis) on channel $A$. In this case, the three parameters required to compensate the phase $\phi$ and perform the Bell inequality measurement are the three phase compensator retardation angles $\chi_{1A}$, $\chi_{2A}$, $\chi_{B}$. 


The device on channel $A$ is represented by the unitary transformation $\hat{T}_A=\hat{T}_{\pi/4}(\chi_{1A})\hat{T}_{0}(\chi_{2A})$, where the corresponding Jones matrices, $\hat{T}_{\pi/4}$ and $\hat{T}_{0}$, are given in \refapp{A:Jones}. The elements of the $\hat{T}$ matrix can then be written as
\begin{equation}\label{Eq:piezo}
\left \{ \begin{array}{l c l}
 h_A&=&\cos\frac{\chi_{2A}}{2}e^{-i\frac{\chi_{1A}}{2} }\\
 v_A&=&- i\sin\frac{\chi_{2A}}{2}e^{i\frac{\chi_{1A}}{2}}.
\end{array}
\right.
\end{equation}
Applying the same procedure as in the previous section, 
we find $\alpha_A = \chi_{2A}/2$, $\varphi_A = -\chi_{1A}/2$, and $\varphi'_A = \chi_{1A}/2- \pi/2$. With the devices considered here on channels $A$ and $B$, the condition $\phi_{\text{eff}} = 0$ (mod $2\pi$) simply gives $\phi_{\text{eff}} = \phi-\chi_{1A}+\pi=0$ (mod $2\pi$).

The phase $\phi$ can be determined periodically in the diagonal basis ($\alpha_A = \alpha_B = \pi/4$, or equivalently $\chi_{2A} = \chi_B = \pi/2$) through the measurement of
\begin{equation}\label{Eq:P++chiA}
\begin{array}{l}
P_{++}^{(\phi)}(\chi_{1A})=\frac{1}{2}\sin^2(\frac{\phi-\chi_{1A}}{2}).
\end{array}
\end{equation}
It can then compensated by the first phase compensator following the relation $\chi_{1A}= \phi + \pi$. Bell measurements can then be performed using the analysis parameters $\alpha_B = \chi_B/2$ and $\alpha_A = \chi_{2A}/2$, as in the case of a linearly polarized Bell state. In this configuration, the development of a fully automated analysis procedure dedicated to elliptically polarized, maximally entangled states can be made simple and rapidly reconfigurable, which would be convenient for future fiber quantum communication systems.

\section{Experimental realization}\label{Sec:exp}

\begin{figure*}
\begin{center}
\includegraphics[width=1.7\columnwidth]{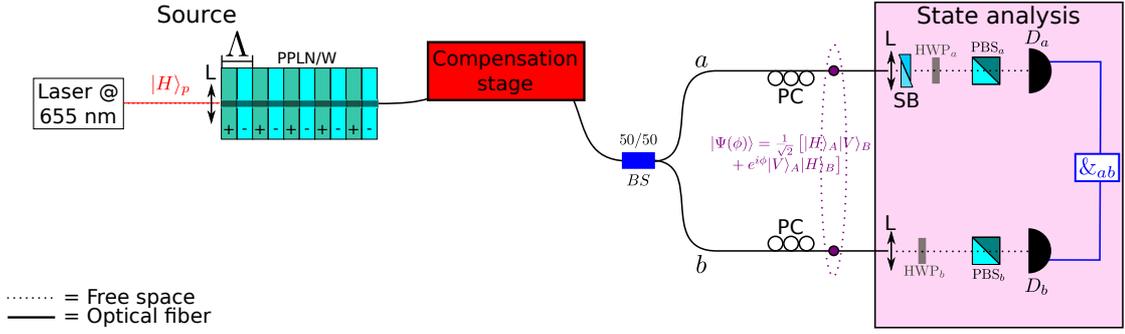}
\caption{\label{fig3} Experimental setup of the considered source and analysis system. Two cross-polarized photons are generated at the degenerate wavelength of 1310\,nm using a type-II periodically poled lithium niobate waveguide (PPLN/W). After being coupled into a telecom fiber, these photons are rendered indistinguishable in terms of spatial and temporal observables thanks to a compensation stage towards providing the state $\ket{H,V}$ (for more details, see the Ref.~\cite{Martin_NJP_2010}). Employing a simple beam-splitter (BS) permits separating the paired photons in a non-deterministic manner, to distribute them to the users, say Alice ($a$) and Bob ($b$), and to create polarization entanglement. In this fibered configuration, a phase between the $\ket{H}$ and $\ket{V}$ components of the entangled state is accumulated along propagation. To cancel this phase and to analyze the entangled state in a proper way, a Soleil-Babinet phase compensator is inserted before the Alice's rotating polarization analyzer. As is commonly the case, Alice and Bob's analyzers are made of an HWP and a PBS.
}
\end{center}
\end{figure*}

In order to generate and experimentally test elliptically polarized maximally entangled states along the lines of the previous analysis, we use the photon-pair source shown in \figurename{~\ref{fig3}}. The source is based on a type-II nonlinear waveguide generator pumped by a 655\,nm laser so as to emit degenerate paired photons at the telecom wavelength of 1310\,nm (see also Ref.~\cite{Martin_NJP_2010} for more details). Instead of the state $\ket{\Phi(\phi)}$ we have been discussing until now, the created photons are in this case in the $\ket{\Psi(\phi)}$ state (see Eq.~(\ref{Eq:Phi})). As mentioned above, the same arguments and state analysis hold for all maximally entangled states.

As the paired photons are distributed over fiber channels, one to Alice ($A$) and the other to Bob ($B$), the ellipticity of the state is due to the birefringence of the fibers, which introduces a phase between the $\ket{H}$ and $\ket{V}$ polarization modes. To perform a Bell inequality measurement, Alice and Bob each have an adjustable linear polarization analyzer consisting of a half wave plate and a polarization beam splitter. Moreover, to compensate the phase, a Soleil-Babinet (SB) device, fixed at $\zeta_{2A}=0$, therefore defined by the transfer matrix of Eq.~(\ref{Eq:Tphizero/4}), is placed on the path of one of the channels, for example on Alice's, as shown in \figurename{~\ref{fig3}}. This setup is slightly different from the situation described in Section~\ref{Sec:DM} as on each channel the rotation parameter of the phase compensator is here replaced by the rotation of a half wave plate (HWP). The free parameters are then the rotation angles $\zeta_{1A}$ and $\zeta_B$ of the half wave plates and the retardation angle $\chi_{2A}$ of the compensator.

\begin{figure}
\begin{center}
\includegraphics[width=0.9\columnwidth]{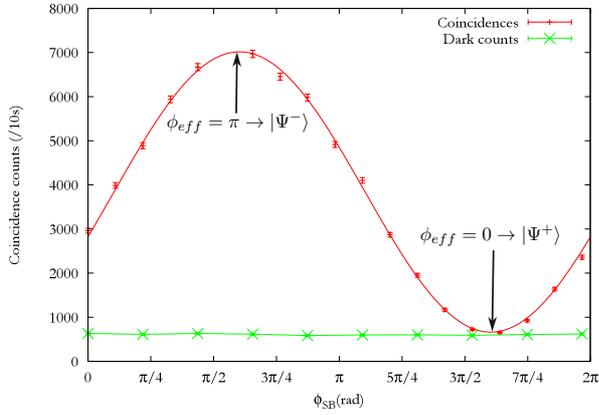}
\caption{\label{fig4}Coincidence rate measured between Alice ($a$) and Bob ($b$) as function of the phase induced by the Soleil-Babinet when the polarization entangled state is analyzed in the diagonal basis, \textit{i.e.}, for both $a$ and $b$ polarization analyzer angles set to $\pi/8$. Note here that $\phi_{SB}$ corresponds $\chi_{2A}$, as defined in Eq.~(\ref{Eq:Tphizero/4}).}
\end{center}
\end{figure}

In order to compensate for the phase of the distributed state, a phase measurement is carried out by setting both HWPs (\textit{i.e.}, $A$ and $B$) at the angle $\zeta_{1A}=\zeta_{B}=\pi/8$ in order to satisfy $\alpha_{A}=\alpha_{B}=\pi/4$. In this configuration, the coincidence probability between detectors $D_A$ and $D_B$ depends on the effective phase of the entangled state (see Eq.~(\ref{Eq:phaseadj})). By tuning the phase $\phi_{SB}$ (corresponding to $\chi_{2A}$) induced by the SB compensator, we observe, as predicted in the case of a maximally entangled state, an interference pattern, shown in \figurename{~\ref{fig4}}. For the specific values of $\phi^1_{SB}\simeq 5$\,rad and $\phi^2_{SB}\simeq 1.86$\,rad for which the coincidence rate is minimized and maximized, we obtain $\phi_{\text{eff}} = 0 $ and $\pi$, corresponding to the maximally entangled states $\ket{\Psi^+}$ and $\ket{\Psi^-}$, respectively. Consequently, using either one of these two phase compensation positions allows maximizing the violation of the Bell inequalities.

\section{Conclusion}\label{C}

In this paper we have addressed that properly measuring and violating the so-called Bell parameter and inequality, respectively, is not always possible when using standard rotating polarization analyzers, since maximally polarization entangled states can either be generated with, or can accumulate along propagation, a phase factor between their two components.

We have developed a general formalism describing how to measure and compensate the phase thanks to the insertion of polarization compensation devices between the source and the polarization analyzers. This approach has also been successfully applied to simple phase correcting schemes, which may present advantages depending on the nature of the experimental setup, \textit{i.e.}, free space or telecom fiber links. We have also linked our theoretical approach to an experimental setup, in which the phase of an entangled state accumulated over fiber channel distribution is canceled by the use of a Soleil-Babinet phase compensator.

Since our approach is general, in the sense that no assumption on the additional polarization devices ($T_A$ and $T_B$) is required, we believe that it can be applicable and useful for various polarization entanglement based quantum communication systems. In particular, this can be of interest when the phase factor in the state is \textit{a priori} not known to the users, which is the case when entanglement is generated from a quantum dot device~\cite{quantum_dot_Mohan,quantum_dot_Beveratos}, or distributed over some distance in optical fibers~\cite{Martin_NJP_2010}.

\section*{Aknowledgements}
The authors thank W. Sohler and H. Herrmann for having provided us with the PPLN/W and for fruitful discussions.

The authors acknowledge the Agence Nationale de la Recherche for the ``e-QUANET'' project (ANR-09-BLAN-0333-01) and the ``FREQUENCY'' project (ANR-09-BLAN-0410), the European ICT-2009.8.0 FET Open program for the ``QUANTIP'' project (grant agreement 244026), the CNRS, the Regional Councils IDF \& PACA, the Institut T\'el\'ecom, the Direction G\'en\'erale de l'Armement (DGA), the Minist\`ere de l'Enseignement Sup\'erieur et de la Recherche, the Direction G\'en\'erale de l'Armement (DGA), the University of Nice -- Sophia Antipolis, and the iXCore Science Foundation, for financial support.

\appendix

\section{Appendix: Jones matrices of a retardation plate}\label{A:Jones}

When polarized light goes through a waveplate, the polarization transformation can be described by the use of the so-called Jones matrices~\cite{jones_new_1941}. This transformation is characterized by the operator $\hat{T}=\hat{R}^{-1}\hat{P}\hat{R}$, $\hat{R}$ and $\hat{P}$ being the rotation and retardation matrices, respectively:
\begin{equation}\label{Eq:RP}
\begin{array}{c c}
\hat{R}=
\begin{pmatrix}
\cos \zeta&\sin \zeta\\
-\sin \zeta&\cos \zeta
\end{pmatrix},
&
\hat{P}=
\begin{pmatrix}
e^{-i\frac{\chi}{2}}&0\\
0&e^{i\frac{\chi}{2}}
\end{pmatrix},
\end{array}
\end{equation}
where $\zeta$ ($-\pi/4\leq\zeta\leq\pi/4$) is the angle between the polarization state $\ket{H}$ and the neutral axis $\ket{H'}$ of the waveplate, and $\chi$ ($-\pi\leq\chi\leq\pi$) corresponds to the retardation phase induced by the anisotropic crystalline plate. The $\hat{T}$ matrix elements (see Eq.~(\ref{Eq:TI})) can then be written:
\begin{equation}\label{Eq:h}
\left \{ \begin{array}{l c l}
 h&=&\cos\frac{\chi}{2}-i\sin\frac{\chi}{2} \cos2\zeta\\
 v&=&-i\sin\frac{\chi}{2}\sin2\zeta.
\end{array}
\right.
\end{equation}

For the specific cases of a half or a quarter waveplate, the corresponding operators can be easily derived as
\begin{equation}\label{Eq:Tlambda/2}
 \hat{T}_{\lambda/2}(\zeta)=-i
\begin{pmatrix}
\cos 2\zeta&\sin 2\zeta\\
\sin 2\zeta&-\cos 2\zeta
\end{pmatrix},
\end{equation}

\begin{equation}\label{Eq:Tlambda/4}
\hat{T}_{\lambda/4}(\zeta)=-\frac{i}{\sqrt{2}}
\begin{pmatrix}
\cos 2\zeta+i&\sin 2\zeta\\
\sin 2\zeta&i-\cos 2\zeta
\end{pmatrix},
\end{equation}
where $\zeta$ is the rotation angle of the considered waveplate. In the specific case of phase compensators at fixed angles $\pi/4$ or $0$, the corresponding operators are given by

\begin{equation}\label{Eq:Tphipi/4}
 \hat{T}_{\pi/4}(\chi)=
\begin{pmatrix}
\cos \frac{\chi}{2} & -i\sin \frac{\chi}{2}\\
-i\sin \frac{\chi}{2} & \cos \frac{\chi}{2}
\end{pmatrix},
\end{equation}

\begin{equation}\label{Eq:Tphizero/4}
\hat{T}_{0}(\chi)=
\begin{pmatrix}
e^{-i\frac{\chi}{2}}&0\\
0&e^{i\frac{\chi}{2}}
\end{pmatrix},
\end{equation}
with $\chi$ the retardation angle of the phase compensator.

\end{document}